\documentstyle[aasms4]{article}

\def\ltsima{$\; \buildrel < \over \sim \;$}
\def\simlt{\lower.5ex\hbox{\ltsima}}
\def\gtsima{$\; \buildrel > \over \sim \;$}
\def\simgt{\lower.5ex\hbox{\gtsima}}

\def\masomen{{_+\atop^{-}}}
\def\ls{{_<\atop^{\sim}}}
\def\gs{{_>\atop^{\sim}}}

\begin{document}

\title{A Long Observation of NGC 5548 by BeppoSAX: the High Energy 
Cut-off, Intrinsic Spectral Variability and a Truly Warm Absorber} 

\author{Fabrizio Nicastro$^{1,2,3}$, Luigi Piro$^{1}$, Alessandra De 
Rosa$^{1}$, Marco Feroci$^{1}$, Paola Grandi$^{1}$, \\
Fabrizio Fiore$^3$, Martin Elvis$^{2}$, Francesco Haardt$^{4}$, 
Jelle Kaastra$^{5}$, Angela Malizia$^6$, Laura Maraschi$^{7}$, 
Giorgio Matt$^{8}$, G. Cesare Perola$^{8}$, 
Pierre Olivier Petrucci$^{7}$}

\affil {$^1$ Istituto di Astrofisica Spaziale - CNR, 
Via del Fosso del Cavaliere, Roma, I-00133 Italy}

\affil {$^2$ Harvard-Smithsonian Center for Astrophysics, 
60 Garden St., Cambridge, MA 02138 USA}

\affil {$^3$ Osservatorio Astronomico di Roma, 
via Osservatorio, Monteporzio-Catone (RM), I00040 Italy}

\affil {$^4$ Dipartimento di Fisica dell'Universit\'a di Milano, 
Milano, Italy}

\affil {$^5$ Space Research Organization Netherlands (SRON), 
Sorbonnelaan 2, 3584 CA Utrecht, The Netherlands}

\affil {$^6$ BeppoSAX Science Data Center, Rome, Italy}

\affil {$^7$ Osservatorio Astronomico di Brera/Merate, 
Milano/Lecco, Italy}

\affil {$^8$ Dipartimento di Fisica, Universit\'a degli Studi ``Roma 3'', 
Via della Vasca Navale 84, I-00146 Roma, Italy}

\today

\begin{abstract}
NGC 5548 was observed by BeppoSAX in a single long (8 day) observation 
from 0.2 to 200 keV. 
We find (1) the spectral variation of the source is produced by a change 
of the intrinsic power law slope; (2) a high energy cut-off at 
$E_c= 115^{+39}_{-27}$ keV with a hint of change of $E_c$ with flux; (3) 
OVII and OVIII absorption K edges, and a possible blended OVII-OVIII 
K$\alpha,\beta$ emission feature at $0.54^{+0.07}_{-0.06}$ keV, 
inconsistent with a purely photoionized gas in equilibrium. 
We propose that the temperature of the absorbing and emitting gas 
is $\sim 10^6$ K so that both collisional ionization and photoionization 
contribute. 

\end{abstract}


\section{Introduction}

NGC~5548 is an extensively UV/X-ray studied, low redshift (z = 0.017), 
bright Seyfert 1 galaxy. 
Evidence for multiple spectral features in the 0.1-20 keV spectrum of 
NGC~5548 abound yet are still controversial: 
a soft excess when in a bright state (a 2-6 keV luminosity of $L_{2-6} 
\gs 2 \times 10^{43}$ erg s$^{-1}$ is the critical value in 
Branduardi-Raymont, 1986; Reynolds, 1997), an ionized absorber 
(Nandra et al., 1993, Reynolds, 1997, George et al., 1998), a Compton 
hump (Matsuoka et al., 1990, Piro et al., 1990, Nandra \& Pounds, 1994), 
and a broad ($\sigma = 0.49^{+0.20}_{-0.14}$ keV) Fe K line at 6.48 
keV (Fabian et al., 1994, Reynolds, 1997). 
This uncertainty is mainly due to the lack of simultaneous broad band 
observations of this variable AGN (Branduardi-Raymont, 1986, Done 
et al., 1995) with sufficient energy resolution and collecting 
area to clearly disentangle the many proposed components. 
BeppoSAX, with its unique array of co-pointed instruments provides a new 
hope of understanding this AGN. Here we present data from a long (8 day) 
observation with BeppoSAX that begins to fulfill these hopes. 

In this paper we present the analysis of a long ($\sim 8$ day) 
BeppoSAX observation of NGC~5548. 
The observation was performed as a part of the BeppoSAX Core 
Program to study the broad-band (0.1--200 keV) spectral variability of 
Seyfert galaxies. 

\section{Data reduction and analysis}

NGC~5548 was observed continuously by BeppoSAX (Boella et al., 1997a) 
for 8 days (1997 August 14-22) with a net exposure of 314 ks. 
We reduced the LECS ({\em Low Energy Concentrator Spectrometer}, Parmar 
et al., 1997), MECS ({\em Medium Energy Concentrator Spectrometer}, Boella 
et al., 1997b) and PDS ({\em Phoswich Detector System}, Frontera et al., 
1997) data following the standard reduction procedures. 
LECS, MECS and PDS data have been screened according to Fiore, Guainazzi 
\& Grandi, 1999, to produce equalized event files. 
For the PDS data we used the fixed rise-time thresholds in the standard 
processing. Data from the 2 MECS units have been merged together 
to increase the signal to noise ratio. 

The average 2-10 keV source intensity of $3.5 \times 10^{-11}$ erg 
cm$^{-2}$ s$^{-1}$ was $\sim 50$ \% lower than that reported by 
Reynolds (1997) and only 6\% lower than the GINGA mean (Nandra \& 
Pounds, 1994). 
This corresponds to a 2-6 keV luminosity of $2.8 \times 10^{43}$ erg 
s$^{-1}$ (using $H_0 = 50$ km s$^{-1}$ and $q_0 = 0.1$ throughout), 
just above the critical value determined by Branduardi-Raymont 
(1986, $L_{2-6} = 2 \times 10^{43}$ erg s$^{-1}$). 
In Fig. 1 a,b,c,d we show the LECS, MECS, PDS and the total 0.1-200 keV 
source lightcurves. 
In the 0.1-3 keV energy range the source brightened during the central 
part of the observation by $\sim 30\%$, while it brightened only 
$\sim 15\%$ in the 2-10 keV band, before returning to the original 
intensity for the last $\sim 100$ ks. 
In contrast the high energy (13-200 keV) lightcurve was consistent with 
being constant over the whole observation. Thus variations 
of $< 15 \%$ were observed when averaged over the entire 0.1-200 keV band. 
In Fig. 1d (bottom panel) we show the MECS 1-3 keV to 3-10 keV hardness 
ratio, which confirms that the source became softer during the 
small flare. 

The long exposure and the long timescale of the observed variability 
($\sim 100$ ks) allowed us to perform detailed spectral and spectral 
variability studies. 
For this purpose we divided the observation into three parts, 
each with  nearly constant intensity and with similar signal to 
noise (Figure 1, Table 1): two low state spectra at the same flux 
level from the first $\sim 120$ ks (L1) and the last 106 ks (L2), 
and a third, high state, spectrum (H) from the central 66 ks 
of the observation. 
We extracted spectra from the three instruments according to 
Fiore, Guainazzi \& Grandi, 1999. 

\section{Model Independent Analysis} 
NGC~5548 experienced strong spectral variability during the BeppoSAX 
observation. 
Fig. 2 (upper panel)  shows this in a model--independent way using 
ratios between the raw data for spectra H and L1. 
The ratio is a smoothly changing function of energy. 
In contrast the L1/L2 (Fig2, lower panel) ratio is consistent with 
no spectral variability between the first and the last part of the 
BeppoSAX observation. 
We note that there are known non-linear effects in the responses 
of the BeppoSAX detectors which could in principle invalidate the 
usefulness of the model-independent analysis based on the ratio of 
the raw data. However the extent of these effects is smaller 
than the amplitude of the features visible in Figure 2. For the 
two imaging detectors LECS and MECS, the non-linearities are 
$\ls 5$\% (MECS: Boella et al., 1997b; LECS: Parmar et al., 
1997). For the PDS these effects are more complex because they involve 
Compton scattering of photons of E$ > 100$ keV in the phoswich 
(Frontera et al., 1997). 
These effects are therefore negligible below this energy for our 
purpose. 

A result similar to that shown in the upper panel of Figure 2 
was found by Done et al. (1995) in ROSAT PSPC data. 
They ascribed the observed variation to a changing soft component. 
However, the smoothness of the H/L1 ratio across three decades in 
X-ray energy makes that explanation unlikely. 
Similarly a change in the opacity or the column of an ionized absorber, 
which could only modify the emerging spectral shape between 
0.5 and $\sim 2$ keV (Nicastro et al., 1999a), is ruled out. 

A simple change in the emitted power-law (F(E) = A E$^{-\alpha}$, with 
A in erg s$^{-1}$ cm${-2}$ $keV^{-1}$, at 1 keV) provides a better fit. 
To better address this point we show, superimposed on the data in 
the upper panel of Fig.2, 3 curves:  (a) a dashed line corresponding 
to a simple 0.1-200 keV flux variation of $+15\%$, (b) a dotted line 
corresponding to a steepening (larger $\alpha$) of spectrum H compared 
to spectrum L1 by $\Delta \alpha = 0.1$ ( H/L1 = 1.36$\times E^{-0.1}$, 
E in keV) with a pivot-point at $\sim 13$ keV and high energy cut-off 
fixed at 115 keV; (c) a solid curve showing a similar change in slope 
($\Delta\alpha = 0.19$, pivot energy at $\sim 6$ keV), plus an 
increase of the {\em e}-folding energy of a high energy cutoff by 
$\Delta E_c = (E_c(L) - E_c(H1)) = 785$ keV ( H/L1 = 1.36$\times 
E^{-0.19} \left( {e^{-E/900} \over e^{-E/115}} \right)$, E in keV). 
These two models are both good representations of the whole 0.1-200 
keV data. 

Finally we note that there is no evidence of change at the energy of the 
iron line in either the H/L1 or L2/L1 ratios, suggesting that the line 
equivalent width and profile do not change during the whole observation. 

\section{The Integrated Low-State Spectrum}

We have shown (\S 3) that NGC~5548 experienced spectral variability 
between L1 and H, but did not vary significantly between the first 
and the last part of the observation (L1 and L2). 
This allows us to sum spectra L1 and L2 to obtain the integrated 
low-state spectrum L = L1+L2 (see Tab. 1). 
The high quality of these data over the full 3 decade range 
of the BeppoSAX instruments allows us to perform a detailed spectral 
analysis on spectrum L, and to discuss in detail each individual 
component. We defer to \S 5 analysis of the spectral variability. 

The integrated BeppoSAX spectrum L is shown in the upper panel 
of Fig.3, along with the best fitting power law model absorbed at 
low energy by cold gas (Model A) constrained 
to have a minimum hydrogen column density equal to the 
Galactic value: N$_H=1.7 \times 10^{20}$ cm$^{-2}$ 
(Stark et al., 1997). 
We allow the relative MECS to PDS normalization to vary by 5\% 
around the value of 0.86, to account for the estimated systematic 
uncertainty (Fiore, Guainazzi \& Grandi, 1999). Analogously we leave 
the LECS to MECS normalization free to vary over the range of 
acceptable values 0.7-1 (Fiore, Guainazzi \& Grandi, 1999). 
Throughout the paper errors are quoted at a confidence level of 90\% for 
1 interesting parameter (i.e. $\Delta\chi^2 = 2.7$). 
The fit is very poor ($\chi^2_r(dof) = 3.06(112)$). 
Five features are clearly visible in the ratio between the data 
and the best fit model (Fig. 3, lower panel): (1) a deficit of 
counts between $\sim 0.6$ and $\sim 1.5$ keV (Nandra et al., 1993); 
(2) an excess below 0.6 keV; (3) a narrow, line-like feature around 
6 keV, at an energy consistent with the K$\alpha$ fluorescence 
line from almost neutral iron (Fabian et al., 1994); (4) a strong 
systematic excess above 10 keV (Matsuoka et al., 1990, Piro et al., 
1990, Nandra \& Pounds, 1994); (5) a marked deficit above 30 keV. 
The first four of these features have already been observed in 
separate narrow band X-ray spectra of NGC~5548, and interpreted 
in models involving the reprocessing of the nuclear radiation by 
both cold and ``warm'' matter close to the nuclear source: 
the low energy excess and deficit of counts can be explained 
by absorption by highly ionized gas obscuring the line of sight to 
the source (Nandra et al., 1993), and the narrow iron line at 6.4 
keV (rest frame) and the high energy bump seen at $\sim 10-20$ keV 
can be described by reflection off optically thick cold matter 
(Nandra \& Pounds, 1994). 

We therefore fitted a model (Model B, Table 2) including these extra 
components:  
a warm absorber (parameterized by the equivalent hydrogen column 
density N$_H$, and the ionization parameter U, defined as the 
ratio of the number density of hydrogen ionizing photons to the 
number density of hydrogen), a Gaussian emission line (to model 
the FeK$\alpha$ emission line at 6.4 keV, rest frame) and a 
reflection hump (with same index as the intrinsic continuum 
and inclination angle $i = 30^o$). 
Note that we do not include an additional soft component. 
Model B greatly improves the fit ($\Delta \chi^2 = 189$, with 
the addition of $\Delta\nu = 6$ free parameters: probability of exceeding 
$F \simeq \Delta \chi^2 / (\Delta \nu \chi^2_r)$ $<< 0.001$), although 
it remains unacceptable ($\chi^2_r(dof) = 1.45(106)$). 

A large deviation above 10 keV is still present (Fig. 4a), along 
with a residual systematic structure between 0.4 and 0.7 keV. 
We added an exponential high energy cut-off to the power law and the 
reflection components, linking the two {\em e}-folding energies to 
the same value (Model C), producing a highly significant improvement 
in the fit (Fig. 4b, $\Delta\chi^2 = 46.9$, with the addition of 1 
free parameter: probability of exceeding F $< 0.001$). The 
fit is now acceptable ($\chi^2_r(dof) = 1.02(105)$). 
However the residuals now show the presence of a narrow $\sim 3 \sigma$ 
excess at $\sim 0.45-0.65$ keV (Fig. 4b). 
The uncertainty on the LECS calibration at 0.6 keV, is less 
than 5\% (Orr et al., 1998), while the deviations between the 
LECS data and Model C at this energy are 10-20\% (Fig. 4b) giving us 
confidence in the reality of this feature. 

\noindent
To model this $\sim 0.45-0.65$ keV excess, we introduced a second, 
narrow ($\sigma = 50$ eV, fixed) Gaussian emission line (Model D), 
which improves the fit by $\Delta\chi^2 = 6.2$ (probability of 
exceeding F of 0.08, with the addition of two parameters). 
The fit is now excellent (Table 3), and the residuals 
are flat over the whole energy range (Fig. 4c). 
The best fit energy of the emission line is consistent with the 
energies of the OVII K$\alpha_{1,2,3}$ triplet transitions 
(E = 0.561, 0.569, 0.574 keV). 

To check the consistency of our findings, we retrieved by the ASCA 
public archive a set of unpublished ASCA observations of NGC~5548 
taken in 1996, and analyzed them to investigate the presence of a 
similar feature in the ASCA data. 
These data clearly confirm the suggestion of a $\sim 0.6$ keV emission 
feature arising from BeppoSAX data. 
We found that a narrow and highly significant excess of counts is left 
at the lower energy end of the ASCA-SIS, after fitting the 0.5-10 keV 
SIS+GIS data with Model B (Costantini, Nicastro \& Elvis , in preparation). 
The energy of these feature, when modelling the data with an additional 
gaussian emission line, is consistent with that found with BeppoSAX. 

Finally we tested a different model for the low energy spectral 
features seen in the BeppoSAX spectrum of NGC~5548. We explored 
the possibility that a high ionized reflector, rather than 
absorber, could account for both the excess of counts below $\sim 
0.6$ keV, and the deficit between $\sim 0.6$ and $\sim 1.5$ keV (Ross \& 
Fabian, 1993). We fitted the 0.1-6 keV LECS+MECS BeppoSAX data with a 
model consisting of an ionized reflector (Magdziarz \& Zdziarski, 1995) 
and a gaussian at 0.54 keV. This model does not produce 
a good fit to the data ($\chi^2_r(dof) = 1.3(73)$), and the residuals 
clearly still show a broad deficit of counts at the energies of the 
OVII and OVIII K-edges. We then rule out this possibility. 

In Table 4 we summarize the components of 
models A-D, along with the statistical results. 

\subsection{Flat Intrinsic Continuum and High Energy Cut-off} 

The BeppoSAX ``low-state'' data of NGC~5548 are well fitted by an 
intrinsic X-ray continuum consisting of a power law that requires 
an exponential cut-off, $E_c = 115^{+39}_{-27}$ keV.
This is the second direct, well-constrained measurement of a high 
energy cut-off in a single AGN (after that of NGC~4151, also measured 
by BeppoSAX: Piro et al, 1998). 
In moderate band width data the power law spectral index and the {\em 
e}-folding energy of the high energy cut-off can be tightly correlated. 
BeppoSAX avoids this degeneracy, as shown in Fig. 5 (dashed lines). 

The intrinsic power law spectral index (measured over the entire 
0.1-200 keV band) is unusually flat ($\Gamma = 1.59^{+0.03}_{-0.02}$). 
Previous measurements (EINSTEIN: Weaver, Arnaud \& Mushotzky, 1995; 
GINGA: Nandra \& Pounds, 1994; ASCA: Reynolds, 1997) 
of the intrinsic 2-20 keV continuum of NGC~5548 gave a steeper 
spectral index ($\Gamma \simeq 1.9-2.0$). 
This discrepancy is not due to the restricted energy 
range of those instruments, compared to BeppoSAX. 
We refitted the low state (L) BeppoSAX data (Model B), using only 
the ASCA SIS+GIS energy range (0.6-8 keV, Table 2, second column), 
and find a power law spectral index that is still flatter than that 
measured with ASCA ($\Gamma_{SAX:0.6-8 keV} = 1.66_{-0.09}^{+0.07}$ 
vs. $\Gamma_{ASCA} = 1.89^{+0.02}_{-0.01}$, Reynolds, 1997), 
implying a real change of slope between the ASCA and BeppoSAX 
observations (taken 4 years apart), comparable with the change 
from L1 to H in the BeppoSAX data. We also note that the $0.6-8$ 
keV spectral index $\Gamma_{SAX:0.6-8 keV} = 1.66_{-0.09}^{+0.07}$, 
is fully consistent with the value measured over the entire 
BeppoSAX band: $\Gamma = 1.59^{+0.03}_{-0.02}$. This does not 
allow us to rule out the presence of a curvature of the nuclear 
X-ray continuum even at energies lower than $\sim 10$ keV, and to 
discriminate (at those energies) between continuum models more 
complicated than a simple power law. 

\subsection{The Warm Absorber and the 0.54 keV Emission Feature} 
To investigate the low energy absorption features we fitted the data 
with a grid of single-zone photoionization models built with CLOUDY 
(V.90.04, Ferland, 1997), for 30 values of the ionization parameter U 
(-1.0 $<$ log~U $<$ 1.9) and 30 values of the equivalent total hydrogen 
column density N$_H$ (21.0 $<$ log~N$_H$ $<$ 22.9). We assume solar 
(Grevesse \& Anders, 1989; Grevesse \& Noels, 1993) abundances of 
the elements from H to Zn. 
These models include only the transmitted spectra. Emission from the 
gas is not considered here. 
We tested the dependence of these models on the electron 
density of the gas $n_e$, and verified that the relative abundance of the 
most abundant ion of the oxygen (OVII) varies by less than 3 \% 
for changes of $n_e$ in the range $10^{-1} - 10^{12}$ cm$^{-3}$ 
(see also Nicastro et al., 1999a). 
We then assumed $n_e = 2 \times 10^9$ cm$^{-3}$, the value needed if 
the ionized absorbing gas was in pressure equilibrium with the gas
of the Broad Emission Line Regions of NGC~5548 (BELRs. See \S 6.3). 
We adopted as an ionizing continuum, the observed NGC~5548 Spectral 
Energy Distribution (SED. Mathur et al., 1995). In the 0.1-150 keV 
energy range we used the cut-off power law measured by BeppoSAX. 
The ionization parameter U is not a good absolute estimator of 
the ionization state of the X-ray absorbing gas. 
A more reliable indicator of the physical ionization state of the 
absorber is the relative ion abundance distribution, which is well 
defined by measuring the relative abundance of at least two abundant 
ions of the most abundant elements. 
We report values of U only for completeness and instead base our 
discussion on the relative ionic abundances in the gas (Table 3). 

Model D gives a column density of the absorber of log~N$_H = 21.44\pm0.12$, 
consistent with both the ASCA estimate of log~N$_H = 21.51^{+0.09}_{-0.13}$ 
[George et et al., 1998: from their best fitting Model B(i)], and the total 
hydrogen column density of the UV absorber, and so consistent with the 
unified X-ray/UV absorber model proposed by Mathur et al. (1995, but see 
also the later results in Mathur et al., 1999). 
OVII and OVIII fractions (Table 3) are also consistent with previous 
X-ray (Reynolds, 1997, George et al., 1998) and UV (Mathur et al., 
1995, 1999; Crenshaw \& Kraemer, 1999) data. 

\medskip
The emission line energy ($0.54^{+0.07}_{-0.06}$ keV) is consistent 
with the oscillator-strength-weighted energy of the strong OVII 
K$\alpha_{1,2,3}$, OVIII$ K\alpha_{1,2}$ and OVII K$\beta$ transitions 
($<E>= 0.60$ keV, Verner et al., 1996, observed at 0.59 keV for the redshift 
of NGC~5548). 
No other obvious candidates are available (e.g. OVI at 0.08 keV, 
NVI at 0.43 keV, NeIX at 0.92 keV: Verner et al., 1996). 
OVII-OVIII are also the most abundant oxygen ions in an photoionized 
gas with log~U=0.7, as in the warm absorber of Model D. 
This makes the ionized absorber gas an excellent candidate 
to produce the detected emission features at 0.54 keV. 
Emission lines from the most abundant ions will inevitably 
be produced in gas in photoionization equilibrium, however their 
intensities and equivalent widths depend strongly on the assumed 
geometry and gas dynamics (Netzer, 1993, 1996; Nicastro, Fiore, 
Matt \& Elvis, 1999c, in preparation). 
In order to test this hypothesis quantitatively we built 
models for ionized absorbers which include the contributions 
of both gas emission and resonant absorption. 
The calculation of the resonant absorption features is carried 
out as described in Nicastro, Fiore \& Matt (1999b), and 
we use CLOUDY (V.90.04, Ferland, 1997) to include the emission 
contribution (whose calculation includes resonant scattering of the 
emission lines). 
A detailed and quantitative presentation of these models is 
deferred to a forthcoming paper (Nicastro, Fiore, Matt \& Elvis, 1999c). 
The equivalent width of the emission lines is much more sensitive than 
the ion relative abundance to the exact value of the electron density. 
We verified with CLOUDY that the equivalent width of the permitted 
OVII K$\alpha$ emission line varies by a factor $\sim 2$ for changes 
of $n_e$ in the range $10^{-1} - 10^{12}$ cm${-3}$, increasing 
monotonically up to densities of $\sim 10^{8}$ cm$^{-3}$, and 
then dropping by $\sim 50 \%$ at $n_e = 10^{12}$ cm$^{-3}$. 
As in our previous pure-absorption models, we assume here $n_e = 2 
\times 10^{9}$ cm$^{-3}$. 
We adopted a spherical geometry and the value of the covering factor 
$f_c$, as seen by the central source, which maximizes the emission 
from the gas. In a static configuration (no bulk motion of the gas) 
this value is $f_c = 0.5$ (Netzer, 1993; Nicastro, Fiore, Matt 
\& Elvis, 1999c), since higher values of $f_c$ would result in larger 
net reabsorption by the near side clouds. 
However if the gas is outflowing/inflowing the emission contribution 
increases monotonically as $f_c$ increases, and reach its maximum for 
$f_c = 1$. 
We also accounted for ``micro-turbulence'' ($\sigma_v$) and bulk 
motion ($v_{bulk}$) of the gas. 
The effect of micro-turbulence is to reduce the opacity at 
the center of the emission lines and to increase the efficiency 
of continuum pumping excitation of their upper levels, and so 
increasing the lines' net intensities. 
The effect of a net bulk motion of the gas is 
that of shifting the absorption lines blueward or redward with 
respect to the corresponding emission lines (and so to increase 
the net equivalent width of the emission lines, if the lines are 
resolved). The tentative identification of the UV absorber 
component associated with the X-ray absorber of NGC~5548 (Mathur et 
al., 1999) gives $v_{bulk} = +500$ km s$^{-1}$, $\sigma_{v} = FWHM / 
(2 \sqrt{ln 2}) = 100$ km s$^{-1}$, which we use here. 
Smaller/larger values of the ratio $(v_{bulk}/\sigma_v)$ would 
result in smaller/larger net equivalent widths of the emission line, 
due to the increasing/decreasing negative contribution of the 
correspondent absorption transition. 
The maximum obtainable value of the net equivalent width of the 
emission lines, however, saturates for $v_{bulk}/\sigma_v \gs 2$ 
(reaching 99\% of its intrinsic maximum for $v_{bulk}/\sigma_v \gs 
6/\sqrt{2} \simeq 4.2$). 

The best fit photoionization equilibrium model (log U=0.70 and 
log~N$_H=21.44$: Model D) predicts three OVII K$\alpha_{1,2,3}$ emission 
lines at 0.561, 0.569, 0.574 keV, along with OVII K$\beta$ at 0.67 
keV and OVIII K$\alpha_{1,2}$ at 0.65 keV (Figure 6, upper panel). 
These lines are the strongest emission features expected between 
0.4 and 0.7 keV (Figure 6, upper panel). 
However the total equivalent width of all six lines (relative to the 
absorbed continuum) is $\ls 8$ eV (at an oscillator-strength-weighted 
energy of $<E>= 0.60$ keV. Figure 6, upper panel), insufficient to explain 
the measured strength of $53^{+41}_{-37}$ eV. 
Since we maximized the emission line strength then, 
unless peculiar and asymmetric geometrical configurations 
are assumed (i.e., our line of sight to NGC~5548 has a lower gas 
column than other directions, and/or the symmetry is conical rather 
than spherical or cylindrical), we can rule out the possibility that 
the same gas in photoionization equilibrium is responsible 
for both the observed absorption and the emission features. 

An alternative physical explanation is that the ionized absorbing 
gas has a higher temperature so that collisional ionization competes 
with photoionization (``hybrid'' model: details are in Nicastro et 
al., 1999a). 
Because in this case the electron temperature lies nearer to the 
line excitation temperature, the emissivity from the single cloud of gas 
is enhanced, thus the emission line equivalent widths are larger, 
relative to pure photoionization equilibrium case (for fixed relative 
abundances of the oxygen ions). 
A model with the same $N_H$ as in the pure-photoionization case (
log N$_H = 21.44$) but the temperature of the gas raised 
to $1.2 \times 10^{6}$ K and log~U = -0.2 is shown in Figure 6 
(lower panel; note the different scale in the vertical axes of the 
two panels). 
With these values the relative abundances of OVII and OVIII in the 
gas are $n_{OVII} = 0.60$ and $n_{OVIII} = 0.35$, similar to those 
obtained with the best fitting model D (Fig. 6, and see Tab. 3). 
The 0.60 keV blend of OVII K$\alpha_{1,2,3}$, OVIII K$\alpha_{1,2}$ and 
OVII K$\beta$ now has EW = 60 eV, seven times as strong as in the pure 
photoionization model, and fully consistent with the observed value 
in NGC~5548.

\subsection{The Reflection Hump and the Iron Line}

The broad band BeppoSAX data allow us to strongly constrain 
the relative amount of reflection, the energy, the width and 
the equivalent width of the iron line, and the primary continuum 
parameters (see \S 4) simultaneously. 

The reflection component is strongly required by our broad band 
BeppoSAX data of NGC~5548. Eliminating this component from Model 
B, and refitting the data, gives $\Delta \chi^2 = -23$ (for one 
interesting parameter, corresponding to a probability of P$< 0.001$). 

The position of the Fe$_{K\alpha}$ line (Tab. 3) is consistent with 
low-to-medium ionized iron (FeI to FeXVI), and its intrinsic 
width is $< 320$ eV at a 99\% confidence level (Tab. 3). 
We measure an equivalent width of EW(Fe$_{K\alpha}$) = $127_{-23}^{+30}$ 
eV, consistent with the measured relative amount of reflection 
$R=0.55_{-0.17}^{+0.19}$. 
Essentially due to the limited energy resolution of the MECS 
at 6 keV, we can not rule out a weak red wing to the iron line, 
although statistically it is not required (a fit with a disk-line 
model with all parameters except the index of the emissivity law and 
the chemical composition free to vary, gives $\chi^2_r(dof)  = 
1.0(100)$, and: $E = 6.34^{+0.35}_{-0.18}$ keV, $6 < R_{in} < 54$, 
$R_{out} > 1800$ --in units of gravitational radii--, $i < 39^o$). 

We checked the consistency of our results with those obtainable 
with narrower band instruments, comparing the 2-10 keV MECS data 
with the 2-10 keV GIS data of a 1996 ASCA observation of NGC~5548. 
In Figure 7 we show the 1, 2 and 3 $\sigma$ confidence levels for 
the two parameters R and $\sigma(Fe)$. Though both the parameters 
are better constrained in the MECS than in the GIS (due to the 
higher signal to noise of the BeppoSAX L spectrum), MECS and GIS data 
are fully consistent with each other. 

%
%

\section{Spectral variability: the ``high-state'' spectrum}

Here we use the well-constrained spectral model D from \S 4 to 
investigate the spectral variability experienced by the source 
during the central part of the observation, H. 

We fitted spectrum H with Model D, again leaving the relative MECS 
to PDS normalization free to vary within the 5\% uncertanty around 
its estimated value of 0.86 (Fiore, Guainazzi \& Grandi, 1999). 
We found that the amount of cold absorption was coincident with 
the Galactic value along the line of sight and that the warm 
absorber and the iron emission line parameters were consistent, 
within the errors, with those found fitting the spectrum L. 
We then fixed the neutral column of gas to the Galactic value,  
and the column density of the warm absorber and the energy of 
the iron emission line to the best fit values of log~N$_H=21.44$ 
and $E = 6.30$ keV found fitting spectrum L, and refitted the 
data. The results are shown in Table 5. For comparison we report 
in the same table the corresponding values for the low-state 
spectrum (same as in table 3). 

All the parameter values are consistent with those obtained 
for spectrum L, except for the slope of the intrinsic continuum 
power law, and the {\em e}-folding energy of the high-energy cut-off. 
The intrinsic continuum power law of spectrum H steepens by 
$\Delta\Gamma = 0.19^{+0.07}_{-0.05}$ compared to spectrum L, 
and the {\em e}-folding energy of the high energy cut-off is  
shifted to higher energy (Figure 5). 

\bigskip
This finding supports the result of \S 3 that a $\Delta\Gamma \sim 0.19$ 
change of the primary power law spectral index, pivoting somewhere in the 
medium X-ray band, can entirely 
account for the observed spectral variability (Fig. 8a). 
As a further check we tested the hypothesis of changes in the 
parameter values of the other components of our model while 
keeping the slope of the intrinsic power law fixed to the 
spectrum L value of $\Gamma = 1.59$. 
We left all the other parameters free to vary. The $\chi^2$ is now 
unacceptably high ($\chi^2_r(dof) = 1.45(107)$) and the residuals 
clearly show the curvature of the intrinsic continuum along with 
many narrow deviations over the whole 0.1-200 keV band (Fig. 8b). 
We also tested the hypothesis of an additional soft variable 
component, which is not visible in spectrum L, but appears in 
spectrum H. We added a blackbody (Case {\em a}) or a second 
power law (Case {\em b}) to Model B, and refitted the 0.1-10 keV part 
of spectra L and H. 
In Case {\em a} we left all the continuum parameters free to vary 
for both spectra. 
In Case {\em b} we forced the slopes of the two power laws to have 
the same values in both the spectra, but left the normalizations free to 
vary between the two state spectra. 
Both these models provide a good description of the data 
($\chi^2_r(dof) = 0.97(188)$ and $\chi^2_r(dof) = 0.97(190)$ for 
Cases {\em a} and {\em b} respectively) but the result is again a 
variation of a single component over the whole 0.1-10 keV band. 
In Case {\em b} the black body normalizations are pegged to zero 
in both spectra, and in Case {\em a} only one of power laws dominates 
the 0.1-10 keV band in both the L and H data, although it is the 
flat one in L and the steep one in H. 
Fixing the ratio of the two normalizations of one of the two 
power laws to the observed value of 1.15 (in the 2-10 keV band), 
gives an unacceptably high $\chi^2 = 1.59(191)$. 
We conclude that the whole 0.1-200 keV power-law continuum 
spectral index of NGC~5548 underwent a significant variation 
during the 300 ks BeppoSAX observation, while the 2-10 keV 
luminosity varied by less than $< 15 \%$. 
This is the first direct evidence of a change of the slope 
of the power law continuum in this Seyfert galaxy. 

\bigskip
We do not detect any change of the ionization state of the 
``warm absorber'', between the two-state spectra L and H. 
However, even accounting for both the changes in ionizing luminosity 
and in spectral shape, and in the best photoionization 
equilibrium case, we would expect changes in $n_{OVII}$ and 
$n_{OVIII}$ of only $+10\%$ and $+3\%$ respectively, well within 
the errors. 

\section{Discussion}

\subsection{Steepening of the Intrinsic Power Law}
The 3 decade broad band of BeppoSAX has allowed us to detect a 
change in the primary continuum shape of NGC~5548, which is 
apparently correlated with slightly different intensity states. 
The slope of the intrinsic power law steepens by $\Delta\Gamma = 
0.19$ as the 0.1-3 keV source flux brightens by a factor of 1.3, 
during the central part of the observation. The 2-10 keV flux 
increases by a factor 1.11 only during the same event. 

Spectral index variations are predicted by both thermal (Haardt, 
Maraschi \& Ghisellini, 1997) and non-thermal (Svensson, 1996) 
models for the production of the X-ray spectrum in Seyfert 1, based 
on variations of the Compton optical depth in the electron and/or 
electron-positron region illuminated by the soft thermal photons 
emitted by the disk. 
However if these regions are pair-dominated (i.e., have large values 
of the compactness parameter), the optical depth variations are 
well correlated with conspicuous (an order of magnitude for $\Delta\Gamma 
= 0.2$, Svensson, 1996), {\em soft} (disk photons) and {\em hard} (X-ray 
photons) luminosity variations (Haardt, Maraschi \& Ghisellini, 1997, 
Svensson, 1996). These are not observed in our observation of NGC~5548. 

Instead we observe $\Delta\Gamma = 0.19$, associated to a small 
0.1-3 keV luminosity change (30\%). 
Fluctuations of the Compton optical depth unrelated to luminosity 
changes are expected in electron-dominated thermal coronae 
(i.e. with small values of the compactness). 
In fact, in this case a relationship between the luminosity and the 
optical depth can not be specified {\it a priori} (Haardt, Maraschi \& 
Gisellini, 1997), and fluctuations of $\tau$, and so of $\Gamma$, 
uncorrelated with luminosity changes are possible. 
Based on our data we can estimate a lower limit for the compactness 
in NGC~5548. The ionizing luminosity of the source in the low-state 
is $L \simeq 2.2 \times 10^{44}$ erg s$^{-1}$, and the timescale 
over which the source change flux by a factor of 2 is greater than 
$\Delta T \sim 8$ days. So the compactness must be $l > (L/c\Delta T) 
(\sigma_T/me_ec^3) \sim 0.2$ (Done \& Fabian, 1989). 
This is a weak limit for the estimated compactness in AGN ($l \sim 
1-100$, Svensson, 1996). 
Our data suggest then that a thermal, electron dominated (low 
compactness) corona is generating the X-ray power law in NGC~5548. 

\subsection{The High Energy Cut-Off}

In the high signal to noise, low-and-flat state spectrum of 
NGC~5548, we clearly measure a steepening of the intrinsic 
spectral shape above $\sim 50$ keV and model the curvature of the 
continuum with a cut-off power law with {\em e}-folding energy of 
$E_c = 115^{+39}_{-27}$ keV. The evidence of a high energy cutoff 
is much weaker in the lower signal to noise, high-and-steep 
state spectrum extracted from the central part of the BeppoSAX 
observation. In this case we measure $E_c > 130$ (at a 
confindence level of $99\%$), which suggests a shift of this 
component to higher energy as the intrinsic power law steepens. 

Both thermal and non-thermal models predict a high 
energy break of the X-ray power law. In non-thermal models 
the steepening of the X-ray power law at high energy is due to 
energy losses by high energy photons which downscatter off 
relatively cold electrons. The energy of the break is then 
of the order of $E_{break} \sim m_e c^2 / \tau_{pair}^2 \sim 5 
l^{-1}$ MeV (where $\tau_{pair}^2$ is the typical number of 
scatterings experienced by a single photon before escaping: 
Svensson, 1996). For $E_{break} \sim 50-170$ keV, $l$ is in 
the range 100-30. 
In thermal models instead an exponential turnover is expected at 
an energy of $E_{c} = 2kT_{corona}$, due to the cut-off of the 
Maxwellian distribution of the electrons in the corona (Haardt, 
Maraschi \& Ghisellini, 1997). Our data do not allow us to 
distinguish between the two different expected spectral shapes. 

Thermal models also predict a strict anticorrelation 
between the temperature of the corona (and hence $E_c$) and the 
steepness of the whole X-ray continuum. Our data would suggest 
the opposite trend. However, we note that the corona--like geometry 
proposed by Haardt (1993), produces an inverse Compton spectrum which 
is not (for E$<<$ kT) a single power--law, but rather resembles a 
broken power--law, with the break depending on T$_{corona}$ (Haardt, 1993, 
Petrucci et al., in preparation). 
A test of more accurate and selfconsistent models for the production 
of the X-ray continuum is deferred to a forthcoming paper (Petrucci et 
al., in preparation). 

NGC~5548 is only the second Seyfert 1 galaxy for which a clear 
detection of a high energy cut-off with relatively low and well 
constrained {\em e}-folding energy (between 50 and 150 keV) has been 
reported. 
The other is NGC~4151 which also shows a very flat X-ray spectrum 
($\Gamma = 1.2-1.5$, Piro et al., 1998). Larger samples of Seyfert 
1s with broad-band X-ray data are needed to find clear answers to 
this issue. 

\subsection{A ``Truly Warm'' Absorber/Emitter}

The ionized absorber of NGC~5548 during the BeppoSAX observation 
has physical and geometrical properties in all respects similar to 
those derived by Mathur et al. (1999), based on their analysis of 
the ASCA and HST GHRS data of NGC~5548. The most abundant ions of 
the oxygen are He-like and H-like. 
We also found a marginal evidence for a sharp emission feature in 
the LECS spectrum, at an energy consistent with that of the 
oscillator-strength-weighted blend of OVII K$\alpha_{1,2,3}$, OVIII 
K$\alpha_{1,2}$ and OVII K$\beta$ emission lines. 
We modelled the emission feature adding a narrow gaussian to our 
model, and obtain an equivalent width of EW=$53_{-37}^{+41}$. 
A similar feature was found in the ASCA spectrum of 
the warm absorber Seyfert 1 galaxy NGC~3783 (George, Turner 
\& Netzer, 1995). 
In that case the column density of the ionized absorber was 
almost an order of magnitude higher than that of the ionized gas 
in NGC~5548. 
The authors concluded that the same photoionized material is 
responsible for the OVII and OVIII absorption and the emission 
line at $\sim 0.6$ keV. 
For NGC~5548 however, if the BeppoSAX detection of the OVII-OVIII 
K$\alpha$, K$\beta$ emission lines is correct (as suggested by the 
1996 ASCA data of this source: Costantini, Nicastro \& Elvis, in 
preparation), the implied 3$\sigma$ lower limit of their total 
equivalent width is EW$> 16$ eV. 
Since (\S 4.2) our pure photoionization model found the maximum 
line strength (for each line) to be $\sim 1.5$ eV, the observed value 
is too large to be produced by the same gas which is responsible 
for the observed OVII and OVIII K edge absorption if photoionization 
equilibrium applies (and a spherical and symmetric geometry is assumed: 
Nicastro, Fiore, Matt \& Elvis, 1999c), even when the blend of OVII 
K$\alpha_{1,2,3}$, K$\beta$ and OVIII K$\alpha_{1,2}$ is considered 
(EW$_{TOT} \ls 8$ eV). 

\noindent
Instead we propose that the gas is kept at a temperature higher 
than the equilibrium photoionization temperature (for example by 
mechanical heating). 
This strongly increases the emissivity from the gas, and would 
simultaneously let photoionization continue driving small and 
rapid changes of the ionization structure (see Nicastro et al., 
1999a, for details). However the relative contibution 
from photoionization and collisional ionization in this gas is 
critical. A gas temperature of $1.2\times 10^6$ K is sufficient 
to produce OVII K$\alpha_{1,2,3}$ emission lines of total equivalent 
width of few tens of eV (with $f_c = 1$), and so 
reproduce the observed quantity. At this temperature, 
if the ionization parameter departed markedly from log~U = -0.2 
(the value used earlier: see \S 4.2, ``hybrid'' model) then the 
ionization structure of the gas would no longer reproduce the 
observed optical depth of the OVII and OVIII K edges. 
To fit the observed features a value of log~U lower than -0.2 must 
be combined with a temperature higher than $1.2\times 10^6$ K, and 
so with a higher value of the gas emission measure. However, this 
produces an excess of soft X-ray emission which is incompatible with 
our data. 
At the opposite extreme, values of log~U higher than -0.2 must 
be accompanied by temperatures $\ls 1.2\times 10^6$ K and so by an 
insufficient gas emissivity to explain the observed OVII-OVIII 
K$\alpha,\beta$ equivalent width. 

The tighteness of this constraints can be used to check the 
consistency between our model of a ``truly warm'' X-ray 
absorber/emitter and the hypothesis of pressure equilibrium 
of the X-ray absorber/emitter of NGC~5548 with the high 
ionization BELRs of this source. 
The condition of pressure equilibrium gives a density 
for the X-ray absorber of: $n_{H_{WA}} = p_{BELR}/T_{WA} \sim 2\times 
10^9$ cm$^{-3}$ ($p_{BELR} \sim 2\times 10^{15}$ K cm$^{-3}$, 
Krolik et al., 1991). 
Hence, using the ionizing luminosity of NGC~5548 as estimated 
by our data (L$_{ion} = 2 \times 10^{44}$ erg s$^{-1}$, which 
gives a rate of ionizing photons of Q$_{ion} = 1.2 \times 10^{54}$ 
ph s$^{-1}$), and the adopted value of log~U = -0.2, we obtain 
a distance of the warm gas from the central source of $R_{WA}
\sim 20$ lt-day, consistent with the estimated distance 
of the BELR (Wandel, Peterson \& Malkan, 1999). 
Finally, assuming a BELR density of $n_{H_{BELR}} = 1-10 \times 
10^{10}$ cm$^{-3}$ (Krolik et al., 1991, Ferland et al., 1992), 
the adopted values of U and $T_{WA}$ imply $U_{BELR} = 0.1-0.01$, 
consistent with the estimates obtained for the BELR of NGC~5548 
based on reverberation studies (Ferland et al., 1992). 
We then propose that NGC~5548 hosts a ``truly warm'' absorber/emitter 
in pressure equilibrium with the BELR clouds, at a distance of 
few lt-day from the central source. Variability studies of this 
component will be crucial to establish the validity of this model. 

\section{Conclusion}

We analyzed the BeppoSAX data of a long observation of the 
Seyfert 1 galaxy NGC~5548, and presented the results of our 
analysis. In the following we summarize our main findings: 

\begin{itemize}

\item{} The broad band of BeppoSAX allowed us to strongly 
constrain the shape of the intrinsic continuum of NGC~5548. 
We found that, for most of the duration of the observation, 
it is well described by an unusually flat, cut-off power law 
with $< \Gamma > = 1.59$, and $E_c = 115$ keV, over the whole 
0.1-200 keV band. 
No additional soft component is required by our data.

\item{} A clear hump around 20-30 keV is present in the raw 
BeppoSAX data of NGC~5548, as well as an apparently narrow 
emission line feature at 6.3 keV. We interpreted both of 
these components as due to reprocessing of the primary 
radiation by cold matter (possibly the accretion disk) 
illuminated by the central source and covering $\sim 50 \%$ 
of the sky seen by the primary radiation, consistent with 
previous observations. 

\item{} We performed time resolved spectral analysis of our 
data and detected a strong, highly energy dependent, spectral 
variability. Most of this spectral variability can be accounted 
for by a simple change of the intrinsic spectral slope, with no, 
or small, total luminosity changes.
We can rule out changes being due to variations in the relative 
intensity of a putative soft component. 

We showed that this behaviour is compatible with the X-ray 
spectrum being generated in a non-pair-dominated hot corona. 

\item{} We confirmed the presence of a warm absorber in NGC~5548, 
with physical properties fully consistent with those deduced by 
previous measurements. OVII and OVIII are also the most abundant 
oxygen ions in the ionized absorber. 

\item{} We found a marginal evidence for a narrow emission feature 
at $\sim 0.6$ keV, consistent with the energy of the 
oscillator-strength-weighted blend of the OVII K$\alpha_{1,2,3}$, 
OVIII K$\alpha_{1,2}$ and OVII K$\beta$ emission lines. 

The 3$\sigma$ lower limit on the equivalent width of this emission 
line is EW$> 16$ eV. 
The same ionized gas, if in photoionization 
equilibrium (and unless peculiar and asymmetric geometries are present), 
cannot be responsible for both the observed 0.5-2 keV absorption 
features and a such a strong feature. 

\item{} We propose, as an alternative, that the ionized absorbing gas 
is truly a warm absorber (at $\sim 10^6$ K), so that collisional 
ionization competes with photoionization. 
The temperature and ionization parameter suggested by our data, 
give a location for the warm gas coincident with the BELR in NGC~5548. 
We also propose that this gas is in pressure equilibrium 
with the BELR clouds of NGC~5548. 

\end{itemize}

\begin{acknowledgements}
F.N would like to thank Smita Mathur for the usefull discussions. 
An anonymous referee provided helpful suggestions which improved 
the presentation of our data. We thank the SAX Scientific Data 
Center. This work was supported in part by NASA grants NAG5-6078 
(LTSA), NAG5-3039 and NAG5-2476. 
\end{acknowledgements}

\newpage


\newpage
%
\figcaption{(a) LECS (0.1-3 keV), (b) MECS (2-10 keV), (c) PDS 
(13-200 keV) and (d) total 0.1-200 keV lightcurves of NGC~5548 during 
the 8-day BeppoSAX observation. 
LECS and MECS data are binned in 5 ks, while for the PDS and the total 
0.1-200 keV band we use a 50 ks binning. 
To highlight the dependence of variability on the particular energy 
range, we used count rate scales with the same ratio in all the four 
panels.
The bottom panel (e) shows the MECS 1-3 keV to 3-10 keV hardness ratio. 
Again data are binned in 5 ks. 
Vertical solid and dashed lines delimit the lower and upper boundaries 
of the time intervals used for extracting the three state spectra L1 
H and L2, as indicated in the second panel.}
%

%
\figcaption{Upper panel: Ratio between the ``high--state'' and the 
``low--state'' spectra H and L1. 
Lower panel: ratio between the two ``low--state'' spectra of NGC~5548, 
L1 and L2. 
The dashed line in the upper panel indicate a $+15\%$ change of the 
0.1-200 keV flux level: the MECS points are distributed around the 
+15\% line. 
The thik solid and dotted lines are the curves (H/L1) = 1.36$\times 
E^{-0.19} \left( {e^{-E/900} \over e^{-E/115}} \right)$ and (H/L1) = 
1.3$\times E^{-0.1}$ (see text for detail).}
%

%
\figcaption{LECS, MECS and PDS L data of NGC~5548, along with the 
best fitting model A (see the text for details). In the lower panel 
we plot the ratio between the data and the best fitting model A.}
%

%
\figcaption{Ratio between the BeppoSAX L data of NGC~5548 and (a) 
the best fitting model B, (b) the best fitting model C, and (c) 
the best fitting model D (see text for details).}
%

%
\figcaption{1, 2 and 3 $\sigma$ confidence levels for the spectral 
index $\Gamma$ and the {\em e}-folding energy $E_c$ for the spectra 
L (dashed lines) and H (solid lines), calculated leaving all the 
parameters of model D free to vary.}
%

%
\figcaption{Emission and absorption by ionized gas in photoionization 
equilibrium with log~N$_H$ = 21.44, log~U=0.70, $v_{bulk} = \sigma_v = 
+500$ km s$^{-1}$, and $f_c = 1$ (upper panel). Lower panel: same 
as the upper panel, but for gas out of thermal photoionization 
equilibrium. The temperature of the gas is kept at $1.2 \times 10^{6}$ 
k, by an external font of heating, and the ionization parameter is 
log~U = -0.2.}
%

%
\figcaption{1, 2 and 3 $\sigma$ confidence levels for the two parameters 
R and $\sigma(Fe)$ for (a) the 2-10 keV MECS data of the BeppoSAX 
observation of NGC~5548 (solid lines), and (b) the 2-10 keV GIS data 
of a 1996 ASCA abservation of NGC~5548 (dashed lines).}
%

%
\figcaption{(a) Ratio between the BeppoSAX data of NGC~5548 during 
the central part of the observation (spectrum H) and the best fitting 
model D. (b) Ratio between spectrum H and the best fitting Model D with 
the slope of the primary power law fixed to $\Gamma = 1.59$.}
%

\newpage
%
\begin{table}[h!]
\caption{Journal of the BeppoSAX observation of NGC 5548}
\normalsize
\begin{tabular}{|c|ccccc|}
\hline
Spectrum & Exposure (ks) & & Count Rate (ct s$^{-1}$) & & S/N \\
& (MECS) & LECS: 0.1-3 keV & MECS: 2-10 keV & PDS: 13-200 keV & (2-10 keV) \\
\hline
L1 & 124 & $0.184\pm 0.002$ & $0.346\pm 0.002$ & $0.83\pm 0.03$ & 200 \\
H & 66 & $0.247\pm 0.004$ & $0.405\pm 0.002$ & $0.86\pm 0.04$ & 165 \\
L2 & 106 & $0.169\pm 0.003$ & $0.349\pm 0.002$ & $0.85\pm 0.04$ & 190 \\
L = L1+L2 & 230 & $0.178\pm 0.002$ & $0.348\pm 0.001$ & $0.84\pm 0.02$ & 280 \\
\hline
\end{tabular}
\end{table}
\begin{table}[h!]
\caption{Spectrum L: best fitting Model B}
\scriptsize
\begin{tabular}{|l|ccc|cc|}
\hline
Energy Range & & Absorption & & \multicolumn{2}{c|}{Continuum} \\ 
\hline
(in keV) & $^a$N$_H^{Cold}$ & $^b$log~N$_H$ & log~U & $\Gamma$ & 
$^c$Norm. \\ 
\hline
0.1-200 & $2.2^{+0.5}_{-0.3}$ & $21.7\masomen0.1$ & $0.95\pm 0.20$ & 
$1.69^{+0.06}_{-0.04}$ & $7.9^{+0.7}_{-0.3}$ \\ 
\hline 
0.6-8 & $1.7^{+3.5}$ & $21.5^{+0.2}_{-0.3}$ & $0.6^{+0.5}_{-0.7}$ & 
$1.66^{+0.07}_{-0.09}$ & $7.4^{+0.7}_{-0.8}$ \\
\hline
\end{tabular}
\begin{tabular}{|l|cccc|c|}
\hline
Energy Range & \multicolumn{4}{c|}{Iron Line and Reflection} & Stat. \\
\hline
(in keV) & $^d$E$_{Fe_{K\alpha}}$ & $^d$$\sigma_{Fe_{K\alpha}}$ & 
$^e$EW(Fe$_{K\alpha}$) & $^f$R & $\chi^2_r(dof)$ \\
\hline
0.1-200 & $6.30^{+0.06}_{-0.05}$ & 
$ < 0.16$ & $140_{-27}^{+37}$ & $0.72^{+0.33}_{-0.37}$ & 1.45(106) \\
\hline
0.6-8 & $6.29\pm0.06$ & $< 0.17$ & 
$120^{+43}_{-30}$ & $ < 1.6$ & 0.91(69) \\ 
\hline
\end{tabular}
\tablenotetext{a}{In $10^{20}$ cm$^{-2}$. The minimum allowed value is 
N$_H = 1.7 \times 10^{20}$ cm$^{-2}$.} 
\tablenotetext{b}{N$_H$ in cm$^{-2}$.}
\tablenotetext{c}{In $10^{-3}$ photons cm$^{-2}$ s$^{-1}$ keV$^{-1}$, 
at 1 keV.}
\tablenotetext{d}{In keV.}
\tablenotetext{e}{In eV.}
\tablenotetext{f}{Relative amount of reflection. The inclination angle is 
fixed at 30$^o$.}
\end{table}
\begin{table}[h!]
\caption{Spectrum L: best fitting Model D}
\scriptsize
\begin{tabular}{|c|ccccc|ccc|}
\hline
& & & Absorption & & & & Continuum & \\
\hline
Model & $^a$N$_H^{Cold}$ & $^b$log~N$_H$ & log~U & $^c$n$_{OVII}$ & 
$^c$n$_{OVIII}$ & $\Gamma$ & $^d$E$_c$ & $^e$Norm. \\
\hline
D & $1.7^{+0.2}$ & $21.44\pm0.12$ & $0.70_{-0.13}^{+0.16}$ & 
$0.63^{+0.09}_{-0.13}$ & $0.31_{-0.08}^{+0.09}$ & 
$1.59_{-0.02}^{+0.03}$ & $115_{-27}^{+39}$ & $7.18_{-0.23}^{+0.27}$ \\
\hline
\end{tabular}
\begin{tabular}{|c|cccc|cc|c|}
\hline
& \multicolumn{4}{c|}{Iron Line and Reflection} & 
\multicolumn{2}{c|}{Oxygen Line or CIP} & Stat. \\
\hline
Model & $^d$E$_{Fe_{K\alpha}}$ & $^d$$\sigma_{Fe_{K\alpha}}$ & 
$^f$EW(Fe$_{K\alpha}$) & R & $^d$(E$_{OVII-OVIII_{K\alpha,\beta}}$ or kT) & 
$^f$EW(OVII-OVIII$_{K\alpha,\beta}$) or $^g$Norm.& $\chi^2_r(dof)$ \\
\hline
D & $6.30^{+0.05}_{-0.06}$ & $ < 0.18$ & $127_{-23}^{+30}$ & 
$0.55_{-0.17}^{+0.19}$ & $0.54_{-0.06}^{+0.07}$ & 
$53_{-37}^{+41}$ & 0.98(103) \\
\hline
\end{tabular}
\tablenotetext{a}{In $10^{20}$ cm$^{-2}$. The minimum allowed value is 
N$_H = 1.7 \times 10^{20}$ cm$^{-2}$.} 
\tablenotetext{b}{N$_H$ in cm$^{-2}$.}
\tablenotetext{c}{$n_{OVII}$ and $n_{OVIII}$ indicate the relative 
abundances of OVII and OVIII ions.} 
\tablenotetext{d}{In keV.}
\tablenotetext{e}{In $10^{-3}$ photons cm$^{-2}$ s$^{-1}$ keV$^{-1}$, 
at 1 keV.}
\tablenotetext{f}{In eV.}
\end{table}
\begin{table}[h!]
\caption{Models A to D}
\scriptsize
\begin{tabular}{|c|cccccc|}
\hline
Model & Cold Absorber & Ionized Absorber & Power Law & Cutoff Power Law & 
Compton Reflection & Fe$_{K\alpha}$: Gauss. Em. Line \\
\hline
A & Y & N & Y & N & N & N \\
B & Y & Y & Y & N & Y & Y \\
C & Y & Y & N & Y & Y & Y \\
D & Y & Y & N & Y & Y & Y \\
\hline
\end{tabular}
\begin{tabular}{|c|cccc|}
\hline
Model & OVII-OVIII$_{K\alpha,\beta}$: Gauss. Em. Line & 
$\chi^2_r(dof)$ & $^a$$\Delta\chi^2$ & $^b$P \\
\hline
A & N & 3.06(112) & & \\
B & N & 1.45(106) & 189 & $<< 0.001$ \\
C & N & 1.02(105) & 46.9 & $< 0.001$ \\
D & Y & 0.98(103) & 6.2 & 0.08 \\
\hline
\end{tabular}
\tablenotetext{a}{With respect to the previous Model.}
\tablenotetext{b}{Probability of Exceeding F.} 
\end{table}
\begin{table}[h!]
\caption{Fit to the spectrum H (Model D):}
\scriptsize
\begin{tabular}{|c|c|ccc|ccc|}
\hline
& Absoprtion & & Continuum & & \multicolumn{3}{c|}{Iron Line and Reflection} \\
\hline
Spectrum & log~U & $\Gamma$ & $^a$E$_c$ & $^b$Norm. & 
$^a$$\sigma_{Fe_{K\alpha}}$ & $^c$EW(Fe$_{K\alpha}$) & R \\ 
\hline
H & $0.70_{-0.30}^{+0.23}$ & $1.78^{+0.03}_{-0.04}$ & 
$> 260$ & $9.8_{-0.4}^{+0.3}$ & $< 0.7$ & $111_{-47}^{+57}$ & 
$0.61_{-0.29}^{+0.32}$ \\ 
\hline
L &  $0.70_{-0.13}^{+0.16}$ & $1.59_{-0.02}^{+0.03}$ & 
$115_{-27}^{+39}$ & $7.18_{-0.23}^{+0.27}$ & $ < 0.18$ & 
$127_{-23}^{+30}$ & $0.55_{-0.17}^{+0.19}$ \\
\hline
\end{tabular}
\begin{tabular}{|c|cc|c|}
\hline
& \multicolumn{2}{c|}{Oxygen Line} & Stat. \\
\hline
Spectrum & $^a$E$_{OVII-OVIII_{K\alpha,\beta}}$ & 
$^c$EW(OVII-OVII$_{K\alpha,\beta}$) & $\chi^2_r(dof)$ \\
\hline
H & $0.58_{-0.26}^{+0.06}$ & $73^{+50}_{-42}$ & 1.07(106) \\
\hline
L & $0.54_{-0.06}^{+0.07}$ & $53_{-37}^{+41}$ & 0.99(102) \\
\hline
\end{tabular}
\tablenotetext{a}{In keV.} 
\tablenotetext{b}{In $10^{-3}$photons cm$^{-2}$ s$^{-1}$ keV$^{-1}$, 
at 1 keV.}
\tablenotetext{c}{In eV.}
\end{table}

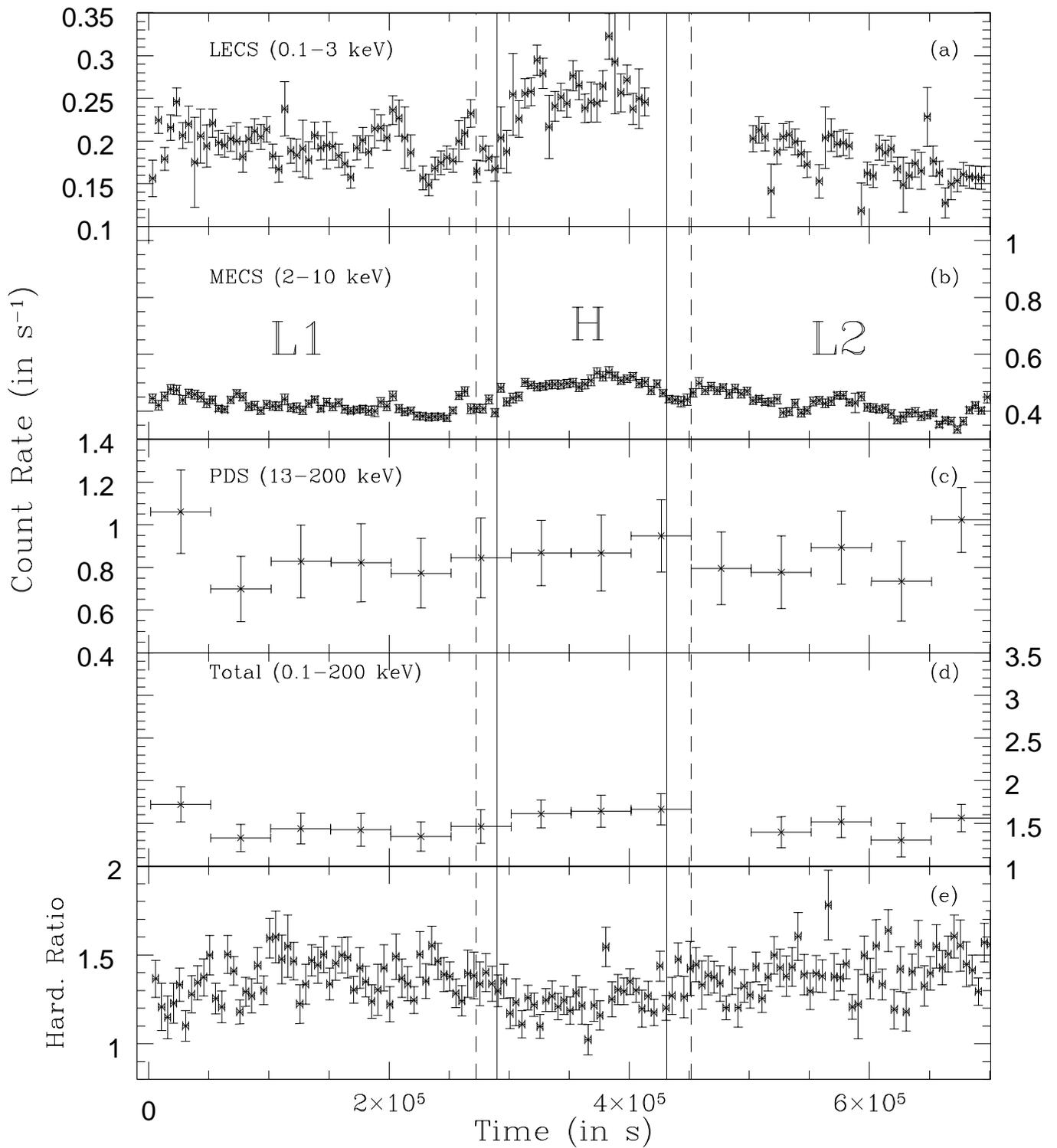
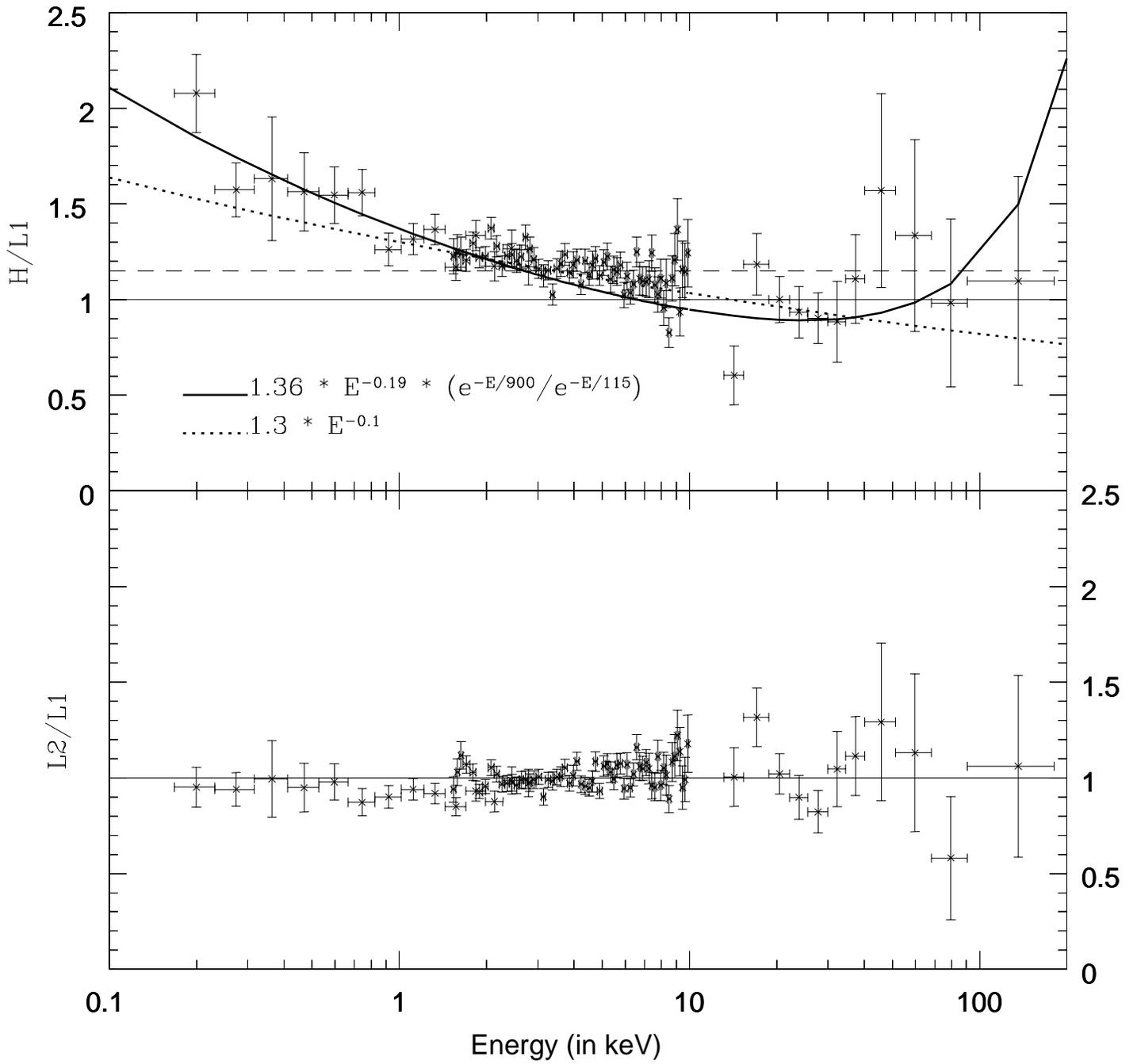
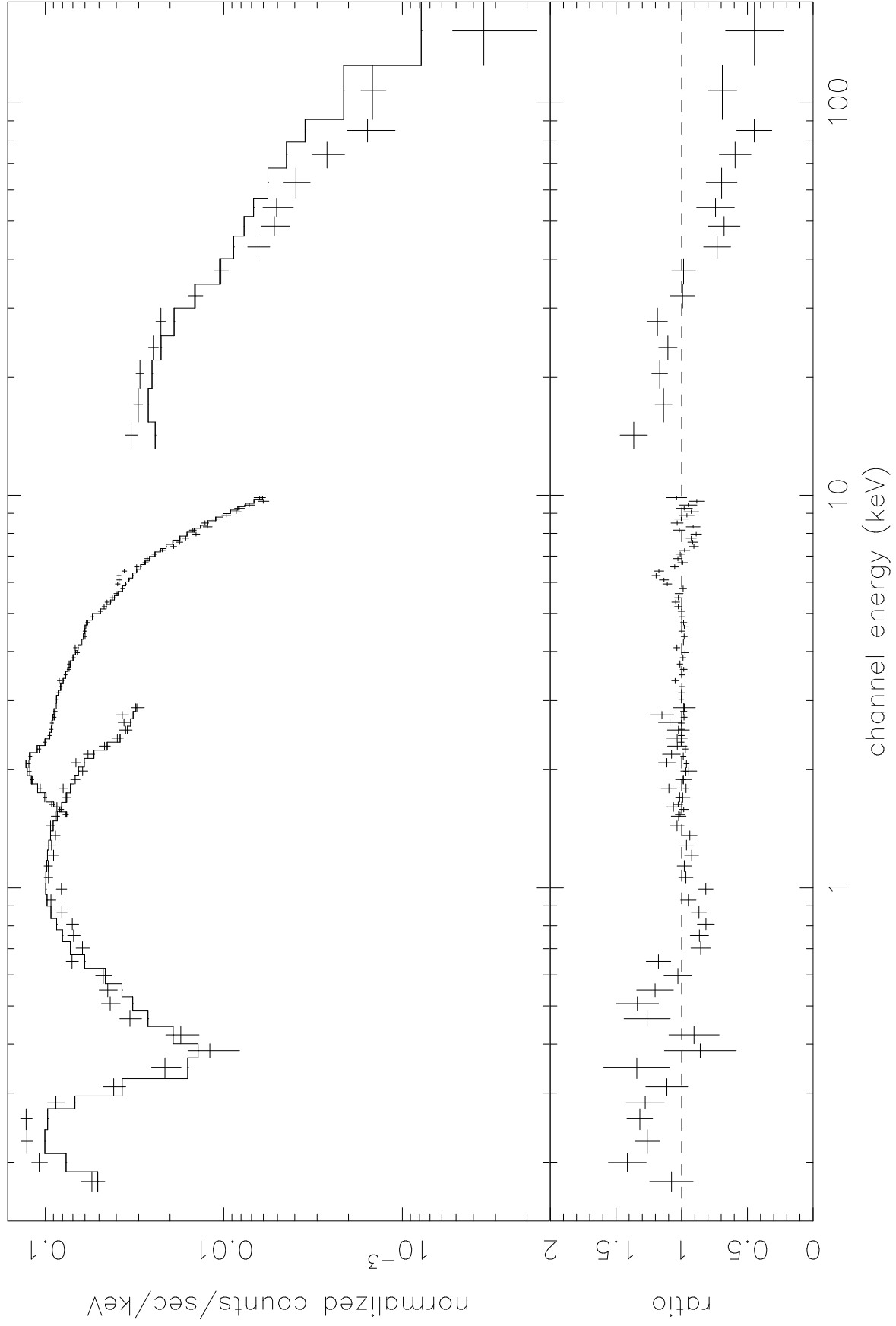
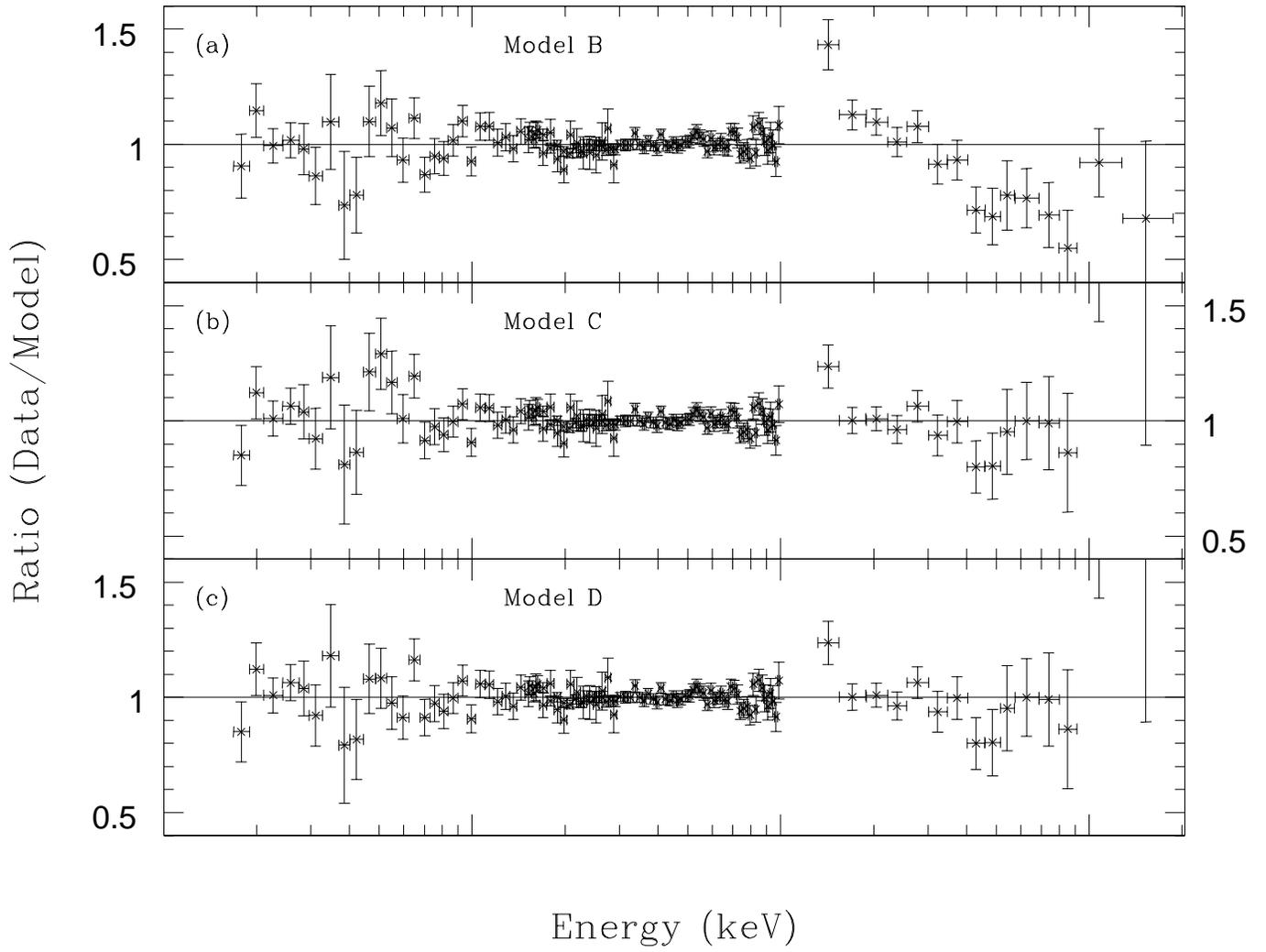
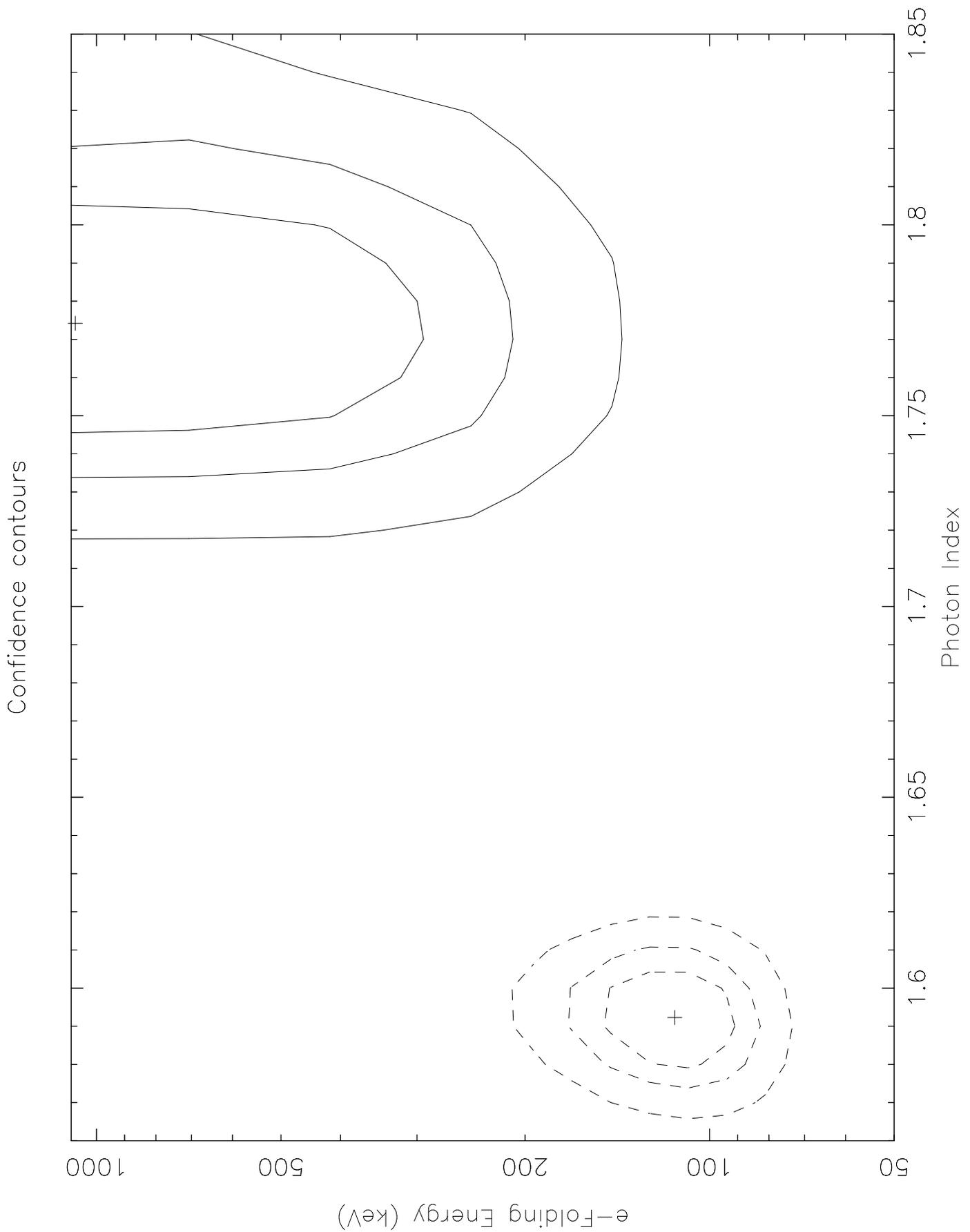
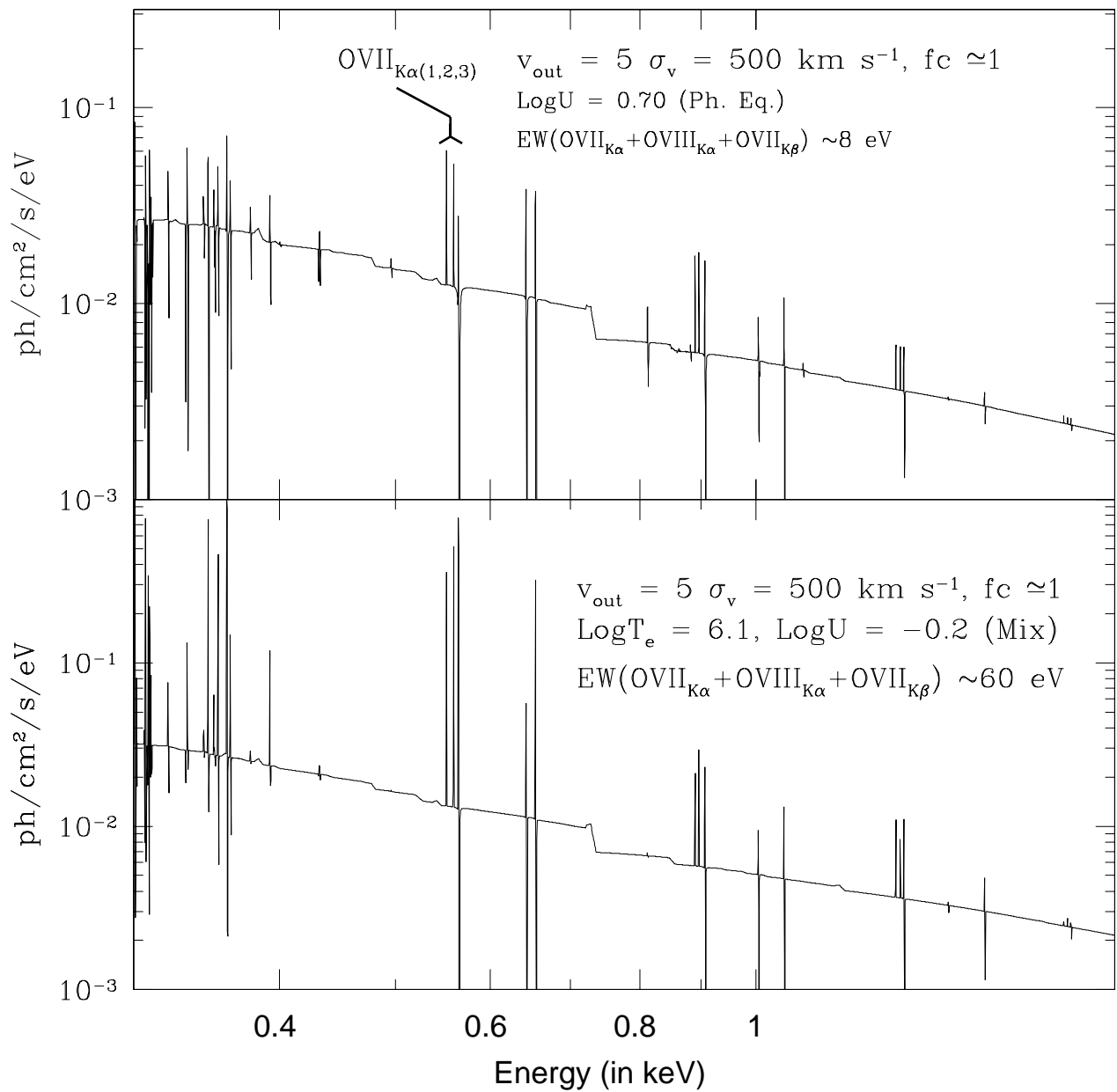
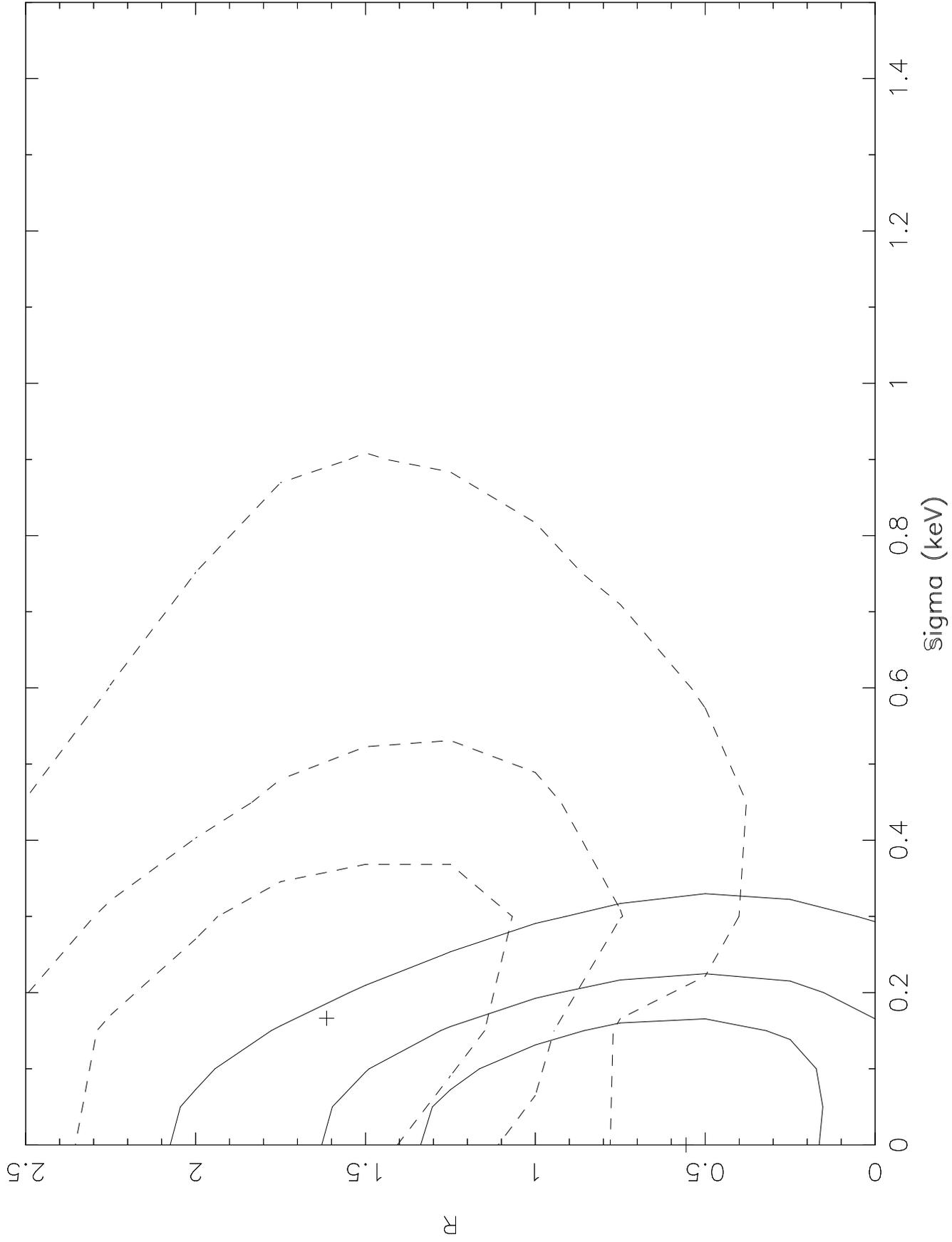
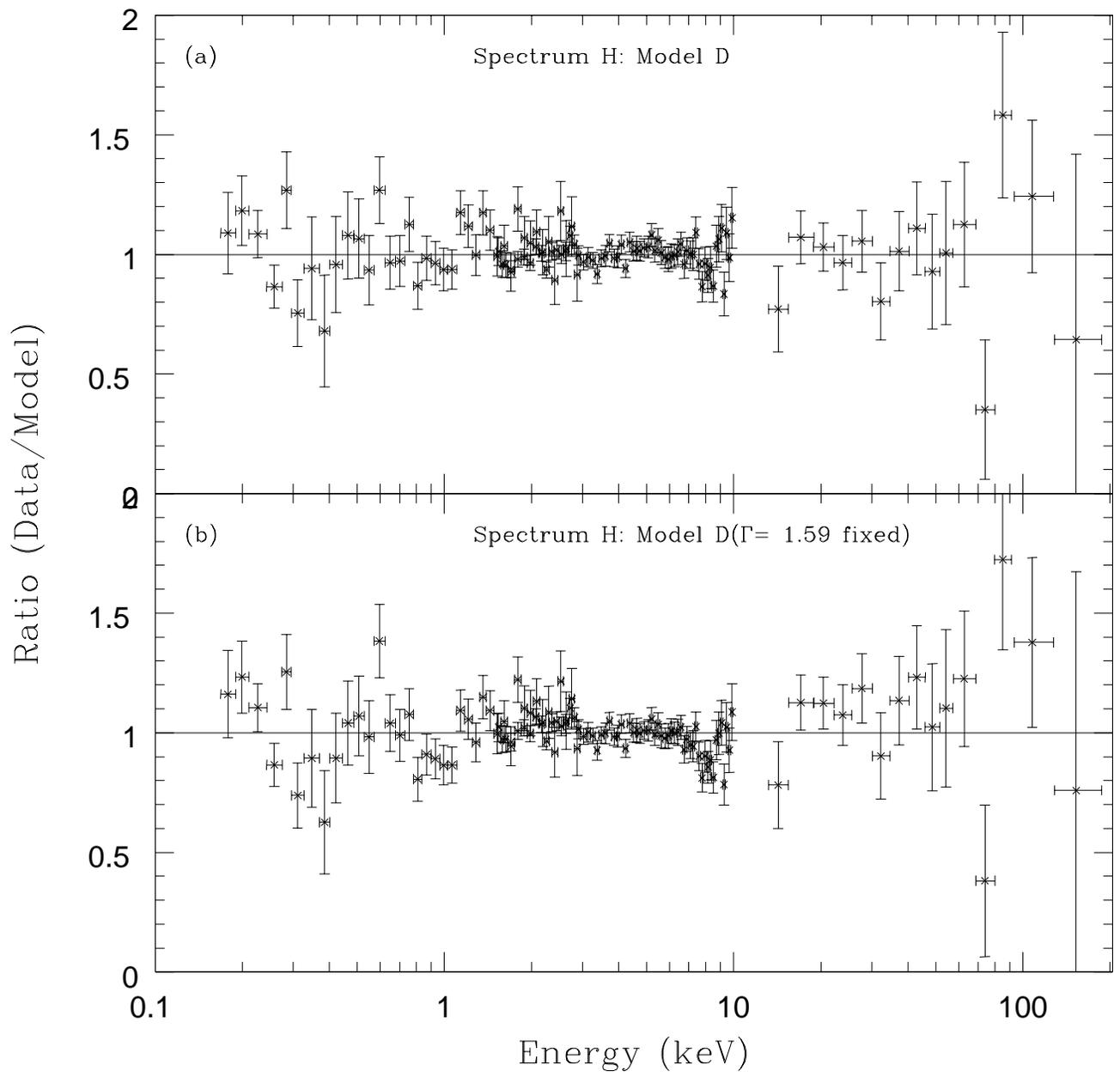
\end{document}